**Wave packet in the phase problem in optics and ptychography.**
Popov N.L., Artyukov I.A., Vinogradov A.V., Protopopov V.V.

1. **Introduction**

The term «optical phase problem» is closely related to the phenomenon of the «lens-less optics». Both cases create a problem, which was first formulated as follows: to reconstruct the phase of the wave field created by a coherent laser beam by measuring its absolute value in two parallel planes: that of the object and that of the detector. An iterative process is used here, one of the components of which is the calculation of the propagation of coherent radiation from the plane of the object towards the plane of the detector. Its result is the distribution of the amplitude of the complex field in both planes. This problem was first considered in [1] with the help of one of the best computers of the time with a memory of ~1 megabyte and a computing power of ~1 megaflop, which allowed analysing a large humber of 32X32 images. The further development of this idea led to the emergence of lensless imaging and ptychography in their current form that is used today in visible, VUV and X-ray ranges [2-9][1]. As the field developed, the original algorithm [1] was modified and made more effective, due in no small part to the evolution of the computing devices used. New algorithms were developed which allowed one to move away from measuring the intensity distribution at the object, instead relying on prior knowledge of it. As a result, lens-less methods of generating images started being used in both experimental and commercial applications [17-20]. One of the main advantages of lens-less optical and ptychographic systems is the possibility of avoiding the problems related to the aberrations of optical imaging devices such as lens, mirrors etc. and approaching the limit of the resolution set by diffraction in microscopy, atmosperic and astronomical optics. Obviously, realizing the aforementioned advantage of lens-less optics would require going beyond the paraxial methods of modeling the propagation of light from the object to the detector. The current work is dedicated to achieving this goal.

2. **Reconstruction of phase (and absolute value) of the coherent field. Wave packet method.**

Most modern phase reconstruction methods are to some degree an evolution of the algorithms described in works made 30 – 40 years ago (see citations in [10,21-23]). These algorithms are given in many articles and continue being improved. They assume that the measurement of the distribution of field intensity at the detector explicitly determines the absolute value of the Fourier transform of the field on the surface of the object within a certain range of spatial frequencies. In practice this is done using the method of calculating the diffraction integral that is most fitting for the experiment.

The algorithms are based on applying four basic operations to the complex target function $f$ multiple times: Fourier transform $F$, replacing its absolute value with an experimentally determined value $A$, inreverse Fourier transform $F^{-1}$ and a transform $P$ for function $f$, which represents the prior knowledge of the object. The result of these transforms is a new function

$$f' = P\left[F^{-1}\left[A\frac{F[f]}{|F[f]|}\right]\right], \tag{1}$$

---

[1] Russian works [10-16] need to be pointed out.

which is given in the same domain as $f$. If $f'$ and $f$ coincide with a certain degree of accuracy, the the objective is accomplished. The *a priory* information ($P$) may be the change of the field's absolute value directly behind the object or in any other plane, the shape of the object, the fact that the field equals zero in the object's plane outside a certain area, the fact that the field is not negative etc.

The layout of a typical lens-less microscope of diffraction-level resolution contains a coherent light source which illuminates the object located in plane $\tilde{S}$ (Fig.1). Unlike the regular microscopes, it does not contain any optical elements between the object and the detector. The laser, illuminating the object from the left, and the computer are not shown.

The importance of the Fourier transform may be shown, using the example wave packet which describes the propagation of coherent radiation from the object $\tilde{S}$ towards the detector $S$ (Fig. 1) and is a precise solution of the wave equation. It determines the field in the half-space through spatial harmonics in the object plane:

$$\psi(\vec{r}) = \iint_{-\infty}^{\infty} \varphi_0(\vec{p}) d^2\vec{p} \exp\{i\vec{p}\vec{\rho} + i\sqrt{k^2 - p^2}z\}, \vec{r} = (x,y,z) = (\vec{\rho},z), \qquad (2)$$

where $k = \frac{2\pi}{\lambda}$ is the wave vector, $\varphi_0(\vec{p})$ – Fourier transform of the field distribution $\psi(\vec{\rho}, z = 0)$ in the object plane. This equation is derived and tied to the other forms of the diffraction integral in [24 – 27]. In the current work, the questions considered will only be related to modeling the propagation of non-paraxial beams relative to reconstructing the phase of the images and ptychography.

Consider that the detector which registers the intensity is located in he far zone, i.e. in the Fourier-plane, where:

$$z \gg \frac{a^2}{\lambda}, \rho = ztg\theta, \qquad (3)$$

Here $a$ is the size of the object, and the aperture angle $\theta$ is assumed to be constant. Considering (2) together with (3) and using the stationary phase method [24], one can determine that the field at the detector takes the form of:

$$\psi(\vec{\rho},z) = \frac{2\pi k}{iz} \frac{z^2}{z^2 + \rho^2} \exp\{ik\sqrt{z^2 + \rho^2}\} \varphi_0\left(\vec{p} = k\frac{\vec{\rho}}{\sqrt{z^2 + \rho^2}}\right). \qquad (4)$$

Equation (4), calculated from the exact solution of the Helmholtz wave equation, shows that the spatial distribution of the field in the far zone is determined by the Fourier transform of the field at the object. In addition, the change in the coordinate $\rho$ at the detector from 0 до $\infty$ leads to the change in the corresponding spatial harmonic $p$ from 0 to $k$. In other words, there is no harmonic in the far zone that would satisfy the condition $p > k$. This means that the maximum spatial resolution in the far zone is determined by the wavelength $\lambda = \frac{2\pi}{k}$ regardless of the optical layout.

In practice, phase reconstruction [21] is performed using a simpler formula – the Fresnel integral which is deribed from (2) with the assumption of small $\theta$:

$$\psi(\vec{\rho},z) = \frac{ke^{ikz}}{2\pi iz} \iint_{-\infty}^{\infty} \psi(\vec{\rho}',0) d^2\vec{\rho}' \exp\left\{ik\frac{(\vec{\rho} - \vec{\rho}')^2}{2z}\right\}, \qquad (5)$$

which in the far zone (3) assumes the form of:

$$\psi(\vec{\rho},z) = \frac{2\pi k}{iz} \exp\left\{ikz\left(1 + \frac{\rho^2}{2z^2}\right)\right\} \varphi_0\left(\vec{p} = k\frac{\vec{\rho}}{z}\right). \qquad (6)$$

Let us compare (6) with the precise value in (4). Comparison of the arguments of the Fourier transform of the starting distribution $\varphi_0(\vec{p})$ shows that the use of the Fresnel integal in the far zone (6) for calculating the field at the detector is possible only if $\rho \leq z$, or $\theta \leq 45°$ and aperture $NA = sin\theta \leq 0.71$. Otherwise one would be dealing with harmonics $p > k$, which, according to (4), are not present in the far zone. Combining this with the requirement that the phase multipliers in (4) and (6) must coincide, leads us to the condition which determines wether or not the Fresnel integral may be used in the far zone (3) [29]:

$$tg\theta \leq 1, \quad \frac{a^2}{\lambda} \ll z \leq \frac{4\lambda}{\pi tg^4\theta}. \tag{7}$$

A more in-depth explanation of the matter is given in [30].

Thus the wave packet approximation in the far zone, just like the paraxial approximation, leads to the aforementioned dependency between the field intensity distribution at the detector and the absolute value of the Fourier transform in the object plane.

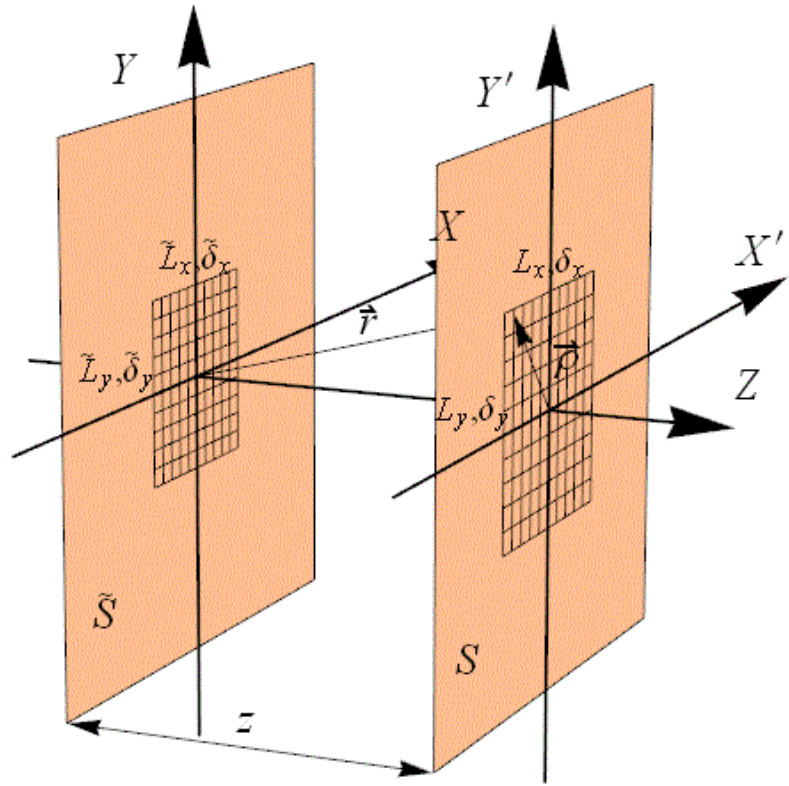

Fig.1. $\tilde{S}$ – object plane, $S$ – detector plane, $z$ – distance between the planes. $(L_x, \delta_x)$ and $(L_y, \delta_y)$ – (size, pixel) detector domains in the $x$ and $y$ directions. $(\tilde{L}_x, \tilde{\delta}_x)$ и $(\tilde{L}_y, \tilde{\delta}_y)$ (size, pixel) objects domains in the $x$ and $y$ directions.

### 3. Discretization of the wave packet method.

From now on, when referring to Fig.1, we will assume the wavelength $\lambda$, detector size $L_x$ and $L_y$, pixel size $\delta_x$ and $\delta_y$ and distance $z$ from the object to the detector to be the set parameters of the experiment. Furthermore, we will set the size of the object domain $\tilde{L}_x$ and $\tilde{L}_y$ (the field of view) and spatial resolution $\tilde{\delta}_x$ and $\tilde{\delta}_y$. This is not the only way to outline the problem, however. For example, one can set the scale and size of the resolved details of the object and z is used to optimize other parameters of the set up.

To make the phase reconstruction algorithms work one must know the absolute value of the Fourier transform, which implies the knowledge of the frequency domain where it is determined. The domain of the desired field in the object plane is inverted relative to the domain in which the absolute value of the Fourier transform is detemined. From chapter 2, one can derive that this frequency domain may be computed from the already known domain in the detector plane:

$$x|y \in \{-L_{x|y}/2 + j \cdot \delta_{x|y}\}, \; j = 0..N_{x|y} - 1,$$
$$p_{x|y} \in \{-\pi/\delta_{x|y} + j \cdot 2\pi/L_{x|y}\}, \; j = 0..N_{x|y} - 1 \quad , \tag{8}$$
$$\delta_{x|y} = L_{x|y}/N_{x|y},$$

where $x|y$ means the direction either along $x$ or along $y$, $p_{x|y}$ – spatial frequency, and $L_{x|y}$, $N_{x|y}$ and $\delta_{x|y}$ – domain size, number and size of the pixels, which correlate to the physical properties of the detector, which we assume to be known. It is also assumed that the coordinate zero is located at the centre (Fig.1).

Choosing the corresponding domain in the object plane is not so easy. We know neither its size, nor the size of its pixels. Its structure can be defined similar to (8), where $\tilde{\delta}_{x|y}, \tilde{L}_{x|y}$ и $\tilde{N}_{x|y}$ – spatial resolution, field of view and the number of pixels on the object. From now on, everything related to the object will be labelled with a wavy line. The link between the domains in the detector plane and the object plane is determined by formula (4). Specifically, the object's spatial frequency $\vec{\tilde{p}}$ is tied to the coordinate at the detector $\vec{\rho}$ by a formula:

$$\vec{\tilde{p}} = k \frac{\vec{\rho}}{\sqrt{z^2 + \rho^2}}. \tag{9}$$

Thus, the maximum value equals:

$$max(\tilde{p}_{x|y}) = k \frac{L_{x|y}/2}{\sqrt{z^2 + (L_{x|y}/2)^2}}. \tag{10}$$

On the oter hand, similar to (8), it equals $\frac{\pi}{\delta_{x|y}}$. This lets us find the optimal (largest) pixel size on the object:

$$\tilde{\delta}_{x|y} = \frac{\pi}{max(\tilde{p}_{x|y})} = \frac{\lambda}{L_{x|y}} z \sqrt{1 + \left(\frac{L_{x|y}}{2z}\right)^2}. \tag{11}$$

The value $\tilde{\delta}_{x|y}$ is the largest object discretization step possible, which still allows its reconstruction. In this case, the aperture of the radiation, leaving a part of the object of size $\tilde{\delta}_{x|y}$, is completely covered by the detector. If the characteristic size of the part of the object is less than this value, then the reconstruction is impossible (with the condition that the object is continuous) and a smaller $z$ must be selected.

In the paraxial approximation $L_{x|y} \ll z$, formula (11) changes into:

$$\tilde{\delta}_{x|y} = \frac{\lambda}{L_{x|y}} z.$$

Thus, the pixel size for the wave packet (11) becomes greater than that in the paraxial approximation and tends to $\lambda/2$ at greater apertures $L_{x|y}/2z \gg 1$.

The object size $\tilde{L}_{x|y}$ must be such that the frequencies $d\tilde{p}_{x|y} = 2\pi/\tilde{L}_{x|y}$, which correspond to the discretization step, are such that the change of coordinate $d_{x|y}$ at the detector is no less than at least one of the $\delta_x$ and $\delta_y$. Otherwise, the detector will not be able to discern frequencies close to each other since they will appear on the same pixel. Assuming $\varphi_0(\vec{p})$ to be sufficiently smooth for $d_{x|y}$ to be determined from the differential, we can use (9) to obtain:

$$\begin{pmatrix} d\tilde{p}_x \\ d\tilde{p}_y \end{pmatrix} = \frac{k}{(\sqrt{z^2 + \rho^2})^3} \begin{pmatrix} z^2 + y^2 & -x\,y \\ -x\,y & z^2 + x^2 \end{pmatrix} \begin{pmatrix} dx \\ dy \end{pmatrix} \tag{12}$$

Accordingly, the reverse differential is:

$$\begin{pmatrix} dx \\ dy \end{pmatrix} = \frac{\sqrt{z^2 + \rho^2}}{k\,z^2} \begin{pmatrix} z^2 + x^2 & x\,y \\ x\,y & z^2 + y^2 \end{pmatrix} \begin{pmatrix} d\tilde{p}_x \\ d\tilde{p}_y \end{pmatrix}. \tag{13}$$

Thus, at least one of the following conditions must hold true:

$$\frac{\sqrt{z^2 + \rho^2}}{z^2}(z^2 + (x|y)^2)\frac{\lambda}{\tilde{L}_{x|y}} \geq \delta_{x|y} \text{ or}$$
$$\frac{\sqrt{z^2 + \rho^2}}{z^2} x\, y\frac{\lambda}{\tilde{L}_{x|y}} \geq \delta_{y|x}. \tag{14}$$

Condition (14) must be fulfilled at all $x$ and $y$, including $x = 0$ and $y = 0$. This means that the domain size in the object plane is:

$$\tilde{L}_{x|y} = \frac{\lambda}{\delta_{x|y}} z. \tag{15}$$

Formulas (11) and (15) determine the spatial resolution and field of view in the object plane. For the sake of convenience6 we will list all the key formulas in the following table.:

| $\tilde{\delta}_{x\|y}$ (11) | $\tilde{N}_{x\|y} = [\tilde{L}_{x\|y}/\tilde{\delta}_{x\|y}]$ | | $\tilde{L}_{x\|y}$ (15) |
|---|---|---|---|
| $\frac{\lambda}{L_{x\|y}}z\sqrt{1 + \left(\frac{L_{x\|y}}{2z}\right)^2}$ | $\left[\left(\frac{L_{x\|y}}{\delta_{x\|y}}\right)/\sqrt{1 + \left(\frac{L_{x\|y}}{2z}\right)^2}\right]$ | $= \left[N_{x\|y}/\sqrt{1 + \left(\frac{L_{x\|y}}{2z}\right)^2}\right]$ | $\frac{\lambda}{\delta_{x\|y}}z$ |

Table.1. Formulas for calculating the spatial resolution $\tilde{\delta}_{x|y}$ and field of view $\tilde{L}_{x|y}$, depending on the detector ($L_{x|y}$, $\delta_{x|y}$), wavelength $\lambda$ and distance $z$. Square brackets mean the closest integer less or equal to the value.

Introducing $NA_{x|y} = \sin\theta_{x|y}$ and $tg\,\theta_{x|y} = L_{x|y}/(2z)$, we get the following classic formula for diffraction resolution from (11):

$$\tilde{\delta}_{x|y} = \lambda/(2NA_{x|y}). \tag{16}$$

Since we are considering the far zone, the condition (3) must hold true, where $a = \tilde{L}_{x|y}$. Accounting for (15), we get from (3):

$$z \leq \frac{\delta_{x|y}^2}{\lambda}. \tag{17}$$

From now on, for the sake of simplicity we will assume a square detector: $\delta_x = \delta_y = \delta$, $L_x = L_y = L$, $N_x = N_y = N$. Then from (11) and (15) follows that the domain in the object plane is also square: $\tilde{\delta}_x = \tilde{\delta}_y = \tilde{\delta}$, $\tilde{L}_x = \tilde{L}_y = \tilde{L}$, $\tilde{N}_x = \tilde{N}_y = \tilde{N}$. According to (15) and (17), a square domain must have a distance

$$z_m = \delta^2/\lambda, \tag{18}$$

at which the field of view $\tilde{L}$ reaches its maximum for the given detector:

$$\tilde{L}_m = \delta. \tag{19}$$

Therefore, placing the detector in the far zone means that the field of view at the object is determined by the detector's pixel size. Substituting $z_m$ into (11), we get the size and number of pixels at the object when the field of view is maximal:

$$\tilde{\delta}_m = \frac{\delta^2}{L}\sqrt{1 + \left(\frac{\lambda L}{2\delta^2}\right)^2}, \tag{20}$$

$$\tilde{N}_m = \frac{\delta}{\tilde{\delta}_m} = \left[\frac{L}{\sqrt{\delta^2 + \left(\frac{\lambda L}{2}\right)^2 \frac{1}{\delta^2}}}\right]. \tag{21}$$

From (21) it follows that the function $\tilde{N}_m(\delta)$ reaches its maximum at $\delta = \delta_M = \sqrt{\frac{\lambda L}{2}}$, so that:

$$\tilde{N}_M = \left[\sqrt{\frac{L}{\lambda}}\right], \tilde{\delta}_M = \frac{\lambda}{\sqrt{2}}, \tilde{L}_M = \sqrt{\frac{\lambda L}{2}} \text{ and } z_M = \frac{L}{2}. \tag{22}$$

This means that, choosing $\delta$ to be equal to $\delta_M$, we maximise $\tilde{N}$ and, respectively, the amount of data acuired. In this case the aperture would obviously be:
$$NA_M = \sin\theta_M = 1/\sqrt{2} \approx 0.71, \theta_M = 45^0.\qquad(23)$$
This aperture allows for the optimum balance between high resolution and large field of view (Fig.2). It should be noted that the maximum number of pixels at the object does not depend on the pixel size at the detector, but is instead determined only by the size of the detector and the wavelength.

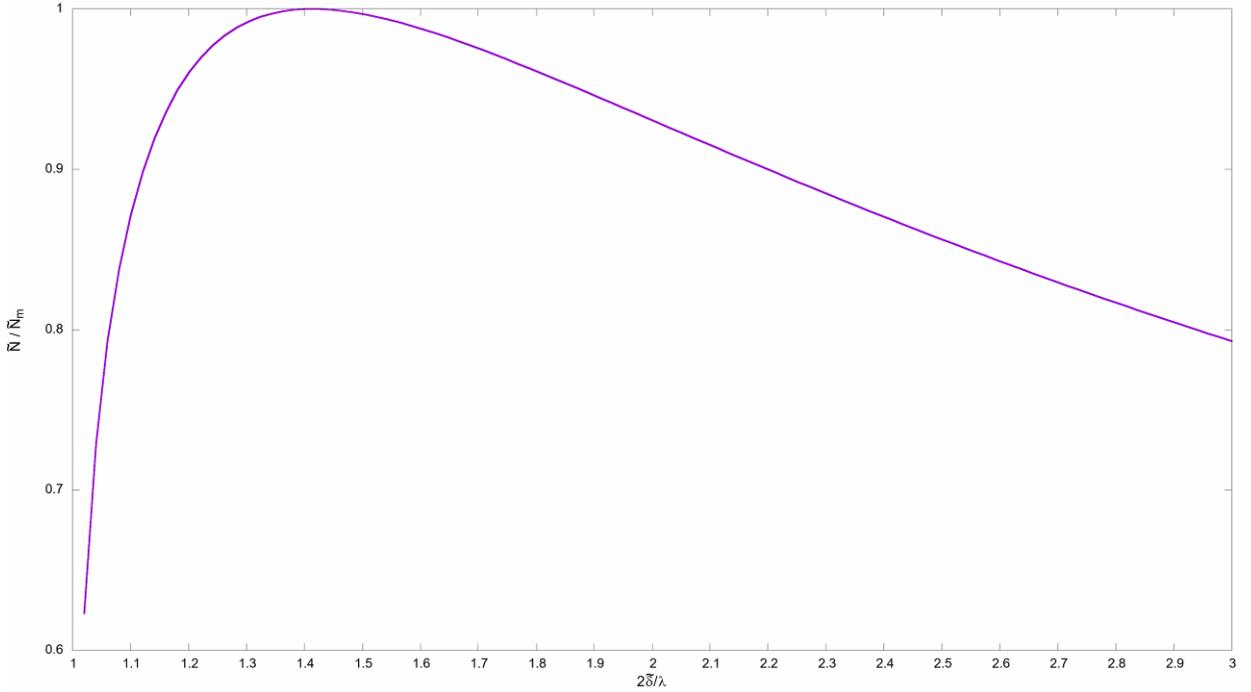

Fig.2. Relationship between $\tilde{N}/\tilde{N}_M$ and $2\tilde{\delta}/\lambda$. The maximum is reached at $\tilde{\delta} = \tilde{\delta}_M = \lambda/\sqrt{2}$.

In practice, the necessary pixel size may be reached by arranging the neighbouring pixels into one virtual pixel. This is called binning – adding up the signal from the neighbouring detector cells. For example, joining four side by side square cells into one doubles the field of view $\tilde{L}_m$, according to (19). At the same time, $z_m$ is quadrupled and the resolution $\tilde{\delta}_m$ is decreased accoring to (20). Aside from increasing the field of view, binning also allows one to increase the dynamic range of the number of received photons per pixel and raises the signal to noise ratio.

From formulas (22) it follows that the largest possible number of pixels at the object $\tilde{N}_M$ rises as $\lambda$ falls. For example, for $\lambda$ = 0.5 μ and $L$ = 1 cm we get $\tilde{N}_M$ = 141, and for $\lambda$ = 10 nm we get $\tilde{N}_M$ = $10^3$. Thus, in order to get resolution in the visible spectrum that is close to the diffraction limit, one must sacrifice the number of pixels at the object and the field fo view related to it $\tilde{L}_M = \tilde{N}_M \times \tilde{\delta}_M$ . However, ptychography (see below) allows one to increase the field of view without losing resolution by using a large number of images.

The above reasoning about the optimal field of view is applicable to the classic lens-less object reconstruction for only one image. In case of the ptychography, the reasoning about the digital domain and optimal object size (19) are applicable to each individual image.

Since the relationship between $\vec{\tilde{p}}$ и $\vec{\rho}$ in (9) is nonlinear, a uniform grid in coordinate system $\vec{\rho}$ is transformed into a nonuniform grid in $\vec{\tilde{p}}$. But in order to use the discrete Fourier trasnform it must be uniform. This requires an interpolation of the function $|\varphi_0(\vec{\tilde{p}}(\vec{\rho}))|$.

Fortunately, in practice, this function is sufficiently smooth, so that interpolation does not pose a problem. However, calculation of the complex amplitude $\psi(\vec{\rho},z)$ using (4), such as in a numeric experiment below, requires a proper interpolation. It may be accomplished by expanding the $\vec{\rho}(\vec{\tilde{p}})$ into the Taylor series around the nearest node. Here the derivatives may be computed using the fast Fourier transform $\partial_{mn}\varphi_0(\vec{\tilde{p}}(\vec{\rho})) = F^{-1}[(i\tilde{p}_x)^m(i\tilde{p}_y)^n F]\varphi_0(\vec{\tilde{p}}(\vec{\rho}))$. For example, achieving the result in the next chapter required calculation of 20 extra discrete Fourier transforms (including 5th order derivatives). It is obvious that the need for interpolation leads to a decreased accuracy and increased calculation time. This is why it is advised to use the paraxial approximation where it is possible, because it does not use interpolation (since $\vec{\tilde{p}} = k\vec{\rho}/z$ is in a linear relationship with $\vec{\rho}$). Obviously, this would lead to the loss of some spatial resolution $\tilde{\delta}$.

4. **Results of computation. Comparison between the wave packet method and the paraxial approximation.**

To evaluate the precision of the wave packet method (4) and the paraxial approximation (6), assuming that the digital domain is computed using (11) and (15), the two methods were compared in the case of the point source function:

$$\psi_s(\vec{r}) = \frac{1}{|\vec{r}|}exp\{ik|\vec{r}|\}$$
$$\varphi_{0s}(\vec{p}) = \frac{i}{2\pi\sqrt{k^2-p^2}} \quad . \tag{24}$$

It is easy to see that substituting (24) into (4) leads to an identity, meaning that the wave packet approximation gives an exact result in the far zone.

For calculations, the wavelength was chosen to be $\lambda$ = 10 nm, while the object and detector plane domains were chosen according to chapter 3. The detector pixel size is $\delta$ = 13 μ, distance and detector size were chosen according to (18) and (22), meaning $z = z_m$ = 1.69 cm, $L = 2z_m$ = 3.38 cm. The resolution $\tilde{\delta}_M$ equals 7 nm in this case.

In Fig.3, one can see the relative error $|(\psi(\vec{r}) - \psi_s(\vec{r}))/\psi_s(\vec{r})|$ of the complex field at the detector, which was computed using the wave packed method (green curve) and the paraxial approximation (blue curve) in comparison to the point source function. The wave packet method gives a relative error of $10^{-16}$ in the interval of angles $tg\theta \in [-1;1]$, with the exception of a small number of angles (clearly artifacts), where the error is $10^{-7}$. The precision of the paraxial method quickly falls as $\theta$ rises, so that at $tg\theta > 0.01$ the relative error is greater than 0.013. According to the applicability estimation for the paraxial method (7), the angle $\theta$ must be sufficiently small: $tg\theta < \sqrt[4]{4\lambda/(\pi z)} = 0.029$. Fig.3 shows that the relative error of the paraxial method, when $tg\theta$ equals this maximum value, is 0.88.

Thus, the numeric experiment with the point source function (24) shows that the error of the paraxial approximation exceeds 50% at the angle of $\theta = 1.66^0$ ($tg\theta = 0.029$) while the result of the wave packet method (4) is almost perfectly precise. The errors apparent on Fig.3 are related to the choice of the domain (see chapter 3) and interpolation. In the range $0 < \theta < 26^0$, the error does not exceed $10^{-16}$ and equals machine precision, while $26^0 < \theta < 45^0$ it is on the same lavel for the vast majority of the points and is around $10^{-7}$ for the rest.

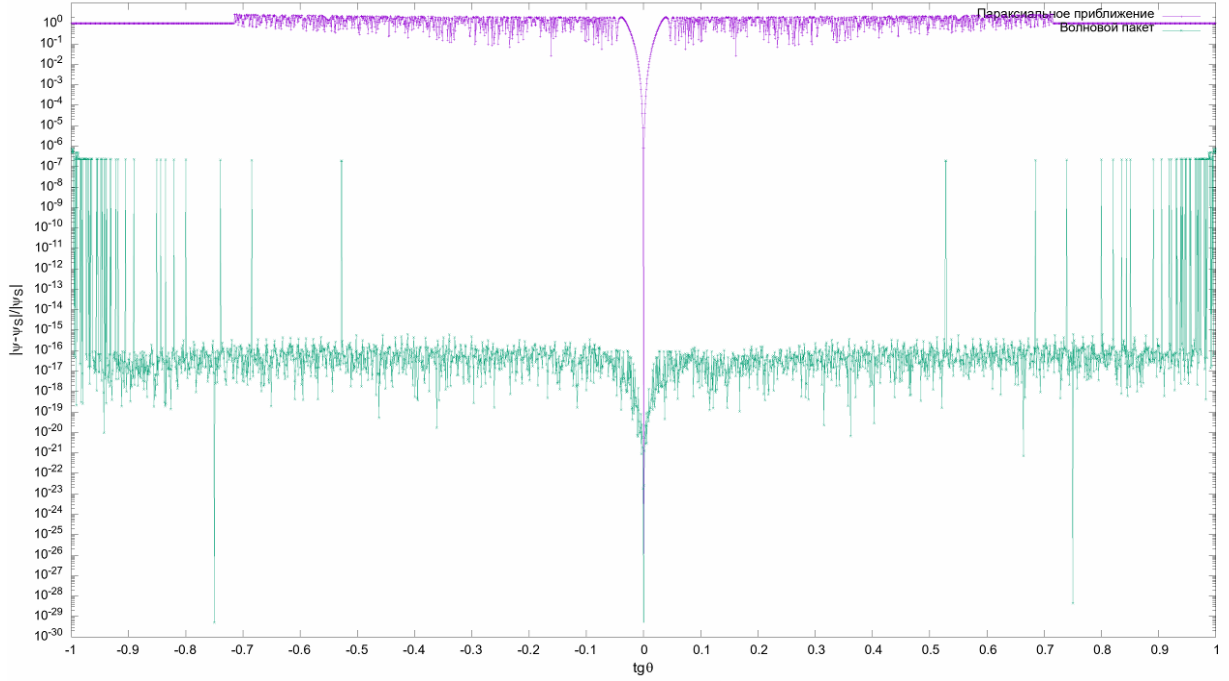

Fig.3. Comparison of the point source solution (24) for $\lambda$=10 nm to the wave packet method (4) (green curve) and the paraxial approximation (6) (blue curve) in the far zone. The X axis is the tangent of the angle with the Z axis; the Y axis is the relative error in logarithmic scale.

### 5. Ptychography. Computation results.

Ptychography is a method of acquiring images using computer processing of overlapping difractograms, which are acquired using small lateral shifts of the observed object (in the $\tilde{S}$ plane on Fig.1) While the object is scanned, the illuminating beam and the detector position remain unchanged. Using several difractograms instead of one (such as with phase reconstruction) allows one to eschew using conditions related to the properties of the studied object. The idea of this method and the term were first published in [31]. In practice, the most used algorithm is ePIE, suggested in [23]. It involves solving a system $J$ equations:

$$A_j(\vec{\tilde{\rho}}) = \left|FFT\left[P(\vec{\tilde{\rho}} - \vec{\tilde{\rho}}_j)O(\vec{\tilde{\rho}})\right]\right|, \quad j = 1...J. \tag{25}$$

Here $P(\vec{\tilde{\rho}} - \vec{\tilde{\rho}}_j)$ – amplitude of the illuminating beam at the surface of the object («illumination function»), $O(\vec{\tilde{\rho}})$ – the desired reflection (or transmission) function of the object, $P(\vec{\tilde{\rho}} - \vec{\tilde{\rho}}_j)O(\vec{\tilde{\rho}})$ – amplitude of the reflected (or passed through) wave, $A_j(\vec{\tilde{\rho}})$ – absolute value of the Fourier transform of the wave, propagating from the object, at its surface (compare to the beginning of chapter 2), $J$ – number of difractograms used in the computations, $j$ – number of the difractograms, $\{\vec{\tilde{\rho}}_j\}_{j=1}^{J}$ – a known set of shifts so that $P(\vec{\tilde{\rho}} - \vec{\tilde{\rho}}_j)$ and $P(\vec{\tilde{\rho}} - \vec{\tilde{\rho}}_{j-1})$ have a 60-70% overlap. The latter is required to guarantee only one solution for (25). The inclusion of the illumination function into consideration is the unique feature of ptychography. It may be known prior to the experiment, determined together with the object, or found by other means. It is important that it must not change during the measurement [23]. More specifically, the algorithm known as extended Ptychography Iterative Engine (ePIE) iterates simultaneous computation of the object function $O(\vec{\tilde{\rho}})$ and the illumination function $P(\vec{\tilde{\rho}})$:

$$O_{j+1}(\vec{\tilde{\rho}}) = O_j(\vec{\tilde{\rho}}) + \alpha \frac{P_j^*(\vec{\tilde{\rho}} - \vec{\tilde{\rho}}_j)}{\left|P_j(\vec{\tilde{\rho}} - \vec{\tilde{\rho}}_j)\right|_{max}^2}\left(f'_j(\vec{\tilde{\rho}}) - f_j(\vec{\tilde{\rho}})\right), \tag{26}$$

$$P_{j+1}(\vec{\tilde{\rho}}) = P_j(\vec{\tilde{\rho}}) + \beta \frac{O_j^*(\vec{\tilde{\rho}} + \vec{\tilde{\rho}}_j)}{|O_j(\vec{\tilde{\rho}} + \vec{\tilde{\rho}}_j)|^2_{max}} \left(f'_j(\vec{\tilde{\rho}}) - f_j(\vec{\tilde{\rho}})\right), \quad (27)$$

$$O_1(\vec{\tilde{\rho}}) = O_0(\vec{\tilde{\rho}}), \quad (28)$$

$$P_1(\vec{\tilde{\rho}}) = P_0(\vec{\tilde{\rho}}), \quad (29)$$

$$f_j(\vec{\tilde{\rho}}) = O_j(\vec{\tilde{\rho}}) P_j(\vec{\tilde{\rho}} - \vec{\tilde{\rho}}_j), \quad (30)$$

$$f'_j(\vec{\tilde{\rho}}) = F^{-1}\left[A_j(\vec{\tilde{\rho}}) \frac{F[f_j(\vec{\tilde{\rho}})]}{|F[f_j(\vec{\tilde{\rho}})]|}\right], \quad (31)$$

$$A_{j+J}(\vec{\tilde{\rho}}) = A_j(\vec{\tilde{\rho}}), \quad \vec{\tilde{\rho}}_{j+J} = \vec{\tilde{\rho}}_j, \quad (32)$$

where $\alpha$ and $\beta$ – dimensionless coefficients of the order of 1, $O_0(\vec{\tilde{\rho}})$ – starting object, $P_0(\vec{\tilde{\rho}})$ – starting illumination, $j$ – iteration number. The difractograms repeat in cycle with the period $J$. The process stops when $O_j(\vec{\tilde{\rho}})$ stops changing. It should be noted that in (26) – (32) the only value that is directly tied with the detector readings is the absolute value of the field amplitude $A_j(\vec{\tilde{\rho}})$. All other variables are determined during computations.

We assumed $\alpha$ = 1, $\beta$ = 0 in our numerical experiment. In this case, the illumination function $P_j(\vec{\tilde{\rho}}) = P_0(\vec{\tilde{\rho}})$ is constant and was chosen to be equal to the point source field located at the range of 10cm inside a round aperture with diameter $\tilde{L}$. Shifts $\vec{\tilde{\rho}}_j$ were chosen to lie on a spiral:

$$\begin{array}{c} \tilde{\rho}_j(\text{мкм}) = \frac{j}{128} 18.2 \left(1 + rand(-0.05, 0.05)\right) \\ \tilde{\varphi}_j = \frac{2\pi}{16} j (1 + rand(-0.05, 0.05)) \end{array}, \quad j = 0..127, \quad (33)$$

where $rand(-0.05, 0.05)$ – random number between -0.05 and 0.05. There is a total of $J$ = 128 overlapping object areas, each 13 μ in diameter. The total field of view is close to a circle 49.4 μ in diameter (Fig.4).

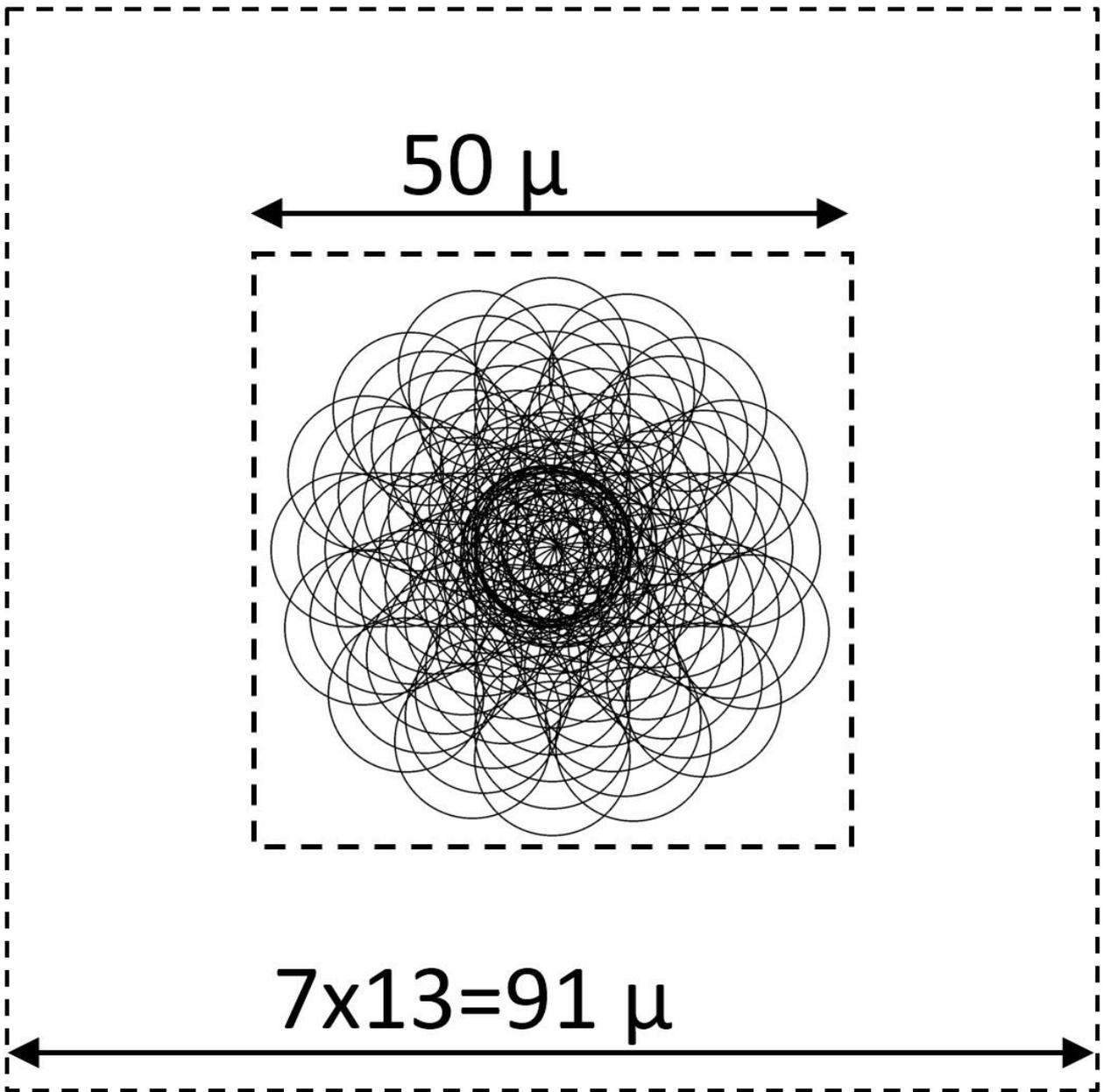

Fig.4. Total field of view composed of 128 circles 13 μ in diameter placed along a spiral. Only the borders of the circles are shown. The total field of view is close to a circle 49.4 μ in diameter and is located inside a square with the side 50 μ. The larger square 91 μ wide depicts the common domain used for calculations.

The digital domain $\tilde{L}$, $\tilde{\delta}$ mentioned in chapter 3 must be used only for calculation (31), since it contains the definition of $A_j(\vec{\rho})$. We will call it the Fourier domain. Since the field of view is 3.8 times larger than the Fourier domain (13x13 μ²), a larger digital domain $7x\tilde{L}$ was selected with the same pixel $\tilde{\delta}$. This is sufficient to fit all the shifts (26) - (30), (33) into the Fourier domain. This domain was used for functions $O_j(\vec{\tilde{\rho}})$, $P_j(\vec{\tilde{\rho}})$, $f_j(\vec{\tilde{\rho}})$ and $f'_j(\vec{\tilde{\rho}})$, where the starting object was set to 1.

    The wavelength was set to 10 nm. The detector was chosen to be square with a square pixel $\delta$ = 13 μ. A fractal template 50x50 μ² wide shown on Fig.5 was used as an object. The height of each number on that template equaled its value, meaning that number «100» was 5 μ high, while «2» was 100 nm high. White colour corresponds to a value of 1, black – to a value of zero.

Let us conduct a preliminary assesment of certain details and the geometry of the experiment. According to (18) and (23), in order to get the maximum amount of information one should have selected $z_m$ = 1.69 cm, $L$ = 3.38 cm, $N$ = 2600. Then, according to (22), the resolution and the field of view would be $\tilde{\delta}_M$ = 7 nm and $\tilde{L}_M$ = 13 μ. However, the physical pixel of a detector measures a discrete value – a number of photons, which cannot be less than zero. This means that the number of photons in one image must be sufficiently large so that there is enough of them for the pixels far away from the axis. The preliminary calculation of the field at the detector using (4) for an object on Fig.5, illumination function $P_0(\vec{\tilde{\rho}})$ and the above geometric parameters $z_m$, $L$ и $N$ = 2600 showed that the number of photons arriving at a pixel near the axis must be ~$10^8$, which is much greater than the saturation threshold of a standard silicon pixel (~$10^3$) at this wavelength. To account for this, we changed the geometry of the experiment by introducing binning of the pixels of a 16x16 detector together with increasing the distance by a factor of 16. According to (15) and (16), size $\tilde{L}$ remained the same (13 μ), while the resolution decreased to $\tilde{\delta}$ = 80 nm. To be able to see a «2» on the reconstructed object we increased the size of the detector, improving the resolution to $\tilde{\delta}$ = 34 nm. In the end, we had $z$ = 27 cm, $L$ = 8 cm, $N$ = 385, $\delta$ = 208 μ, $\tilde{L}$ = 13 μ, $\tilde{\delta}$ = 34 nm. The resulting photon distribution at the detector is shown on Fig.6. The maximum value equals $1.75 \cdot 10^6$, which corresponds to ~7000 photons per pixel.

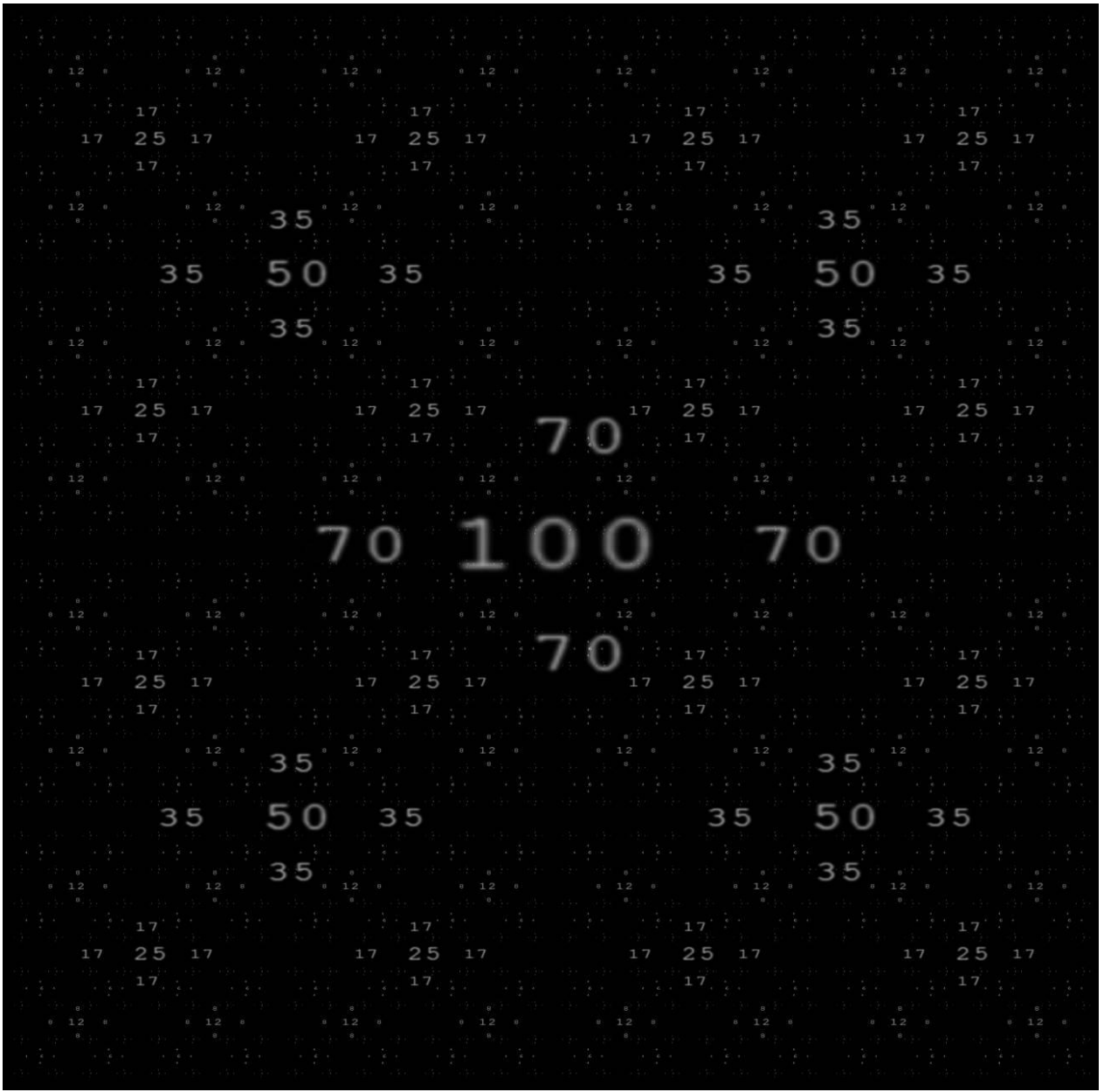

Fig.5. Object in the form of a fractal template. Size is 50x50 µ². The height of the number equals «number value» * 50 nm. I.e., the height of the number «100» is 5000 nm, the height of the smallest number «2» is 100 nm. White colour corresponds to the value 1, black – to the value 0.

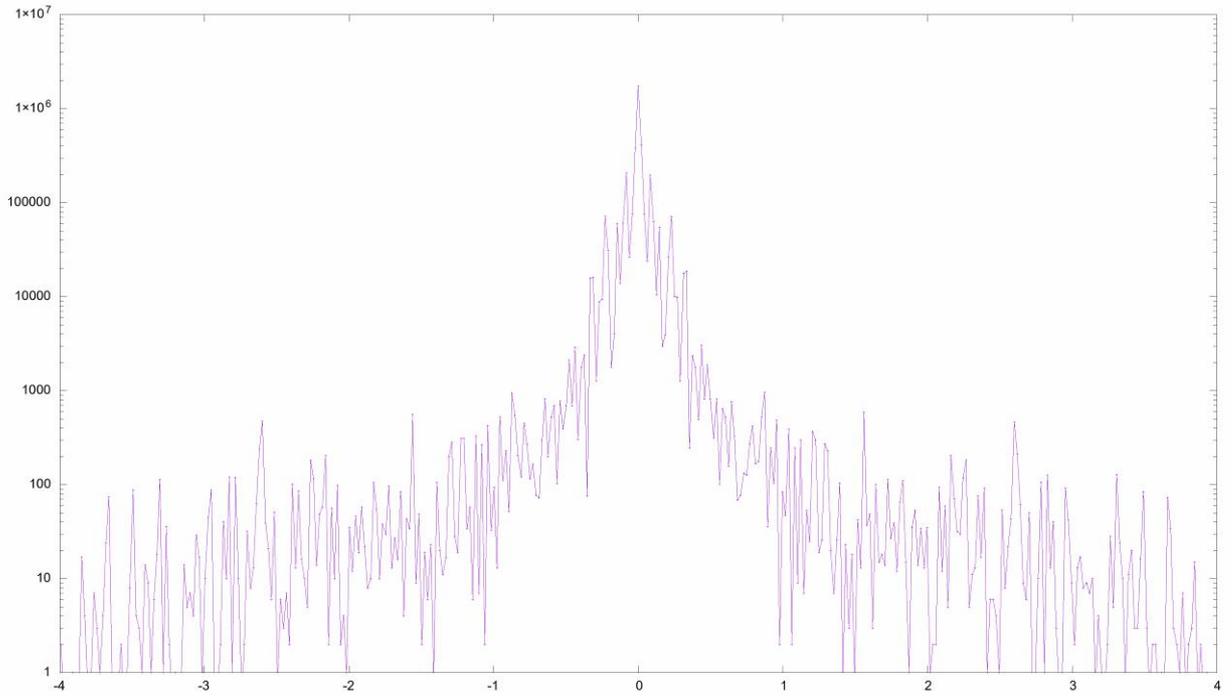
Fig.6. Photon distribution along the detector.

At the first stage, the difractograms of the field $O(\vec{\tilde{\rho}})P_0(\vec{\tilde{\rho}} - \vec{\tilde{\rho}}_j)$ were calculated using (4) for all 128 shifts $\{\vec{\tilde{\rho}}_j\}$ (33). The illumination field $P_0(\vec{\tilde{\rho}})$ was normalized so that the square of its absolute value equaled the number of photons hitting a unit of surface, so that the integral of the field intensity across the pixel at the detector gave the number of photons that hit it. After that, we performed a rounding to the nearest integer and introduced noise according to Poisson's distribution. Following that, the formula (4) was used to reverse-compute the absolute value of the Fourier transform of the object $A_j(\vec{\tilde{\rho}})$ relative to the distribution of photons at the detector (Fig.6). A simple zero order interpolation was used for this.

At the second stage, we applied a ptychography algorithm (26) – (32). After the 10112$^{nd}$ iteration (79 cycles across 128 ptychograms), the reconstructed object stopped changing and took the form shown on Fig.7. The white areas on the periferal sides of the image is the remainder of $O_0(\vec{\tilde{\rho}})$ = 1 which was not touched by the algorithm, since the size of the field of view was 49.4x49.4 μ$^2$ , which is less than the size of the object 50x50 μ$^2$. Fig.8 shows the comparison between the central parts of the reconstructed (left) and original (right) image of 5x5 μ$^2$ in size. It is possible to find the number «2» on the reconstructed image, but impossible to discern it. This corresponds to the declared resolution of 34 nm, meaning that the font «2» takes up 3x3 pixels, which is insufficient for identification, but sufficient for spotting.

The process of reconstructing the image from the first iteration to the 10112$^{nd}$ one took 2 hours, 21 minutes or 0.84 seconds per iteration. Computations were performed on a 14 core CPU Intel(R) Core(TM) i9-7940X CPU @ 3.10GHz and 128 Gb of RAM.

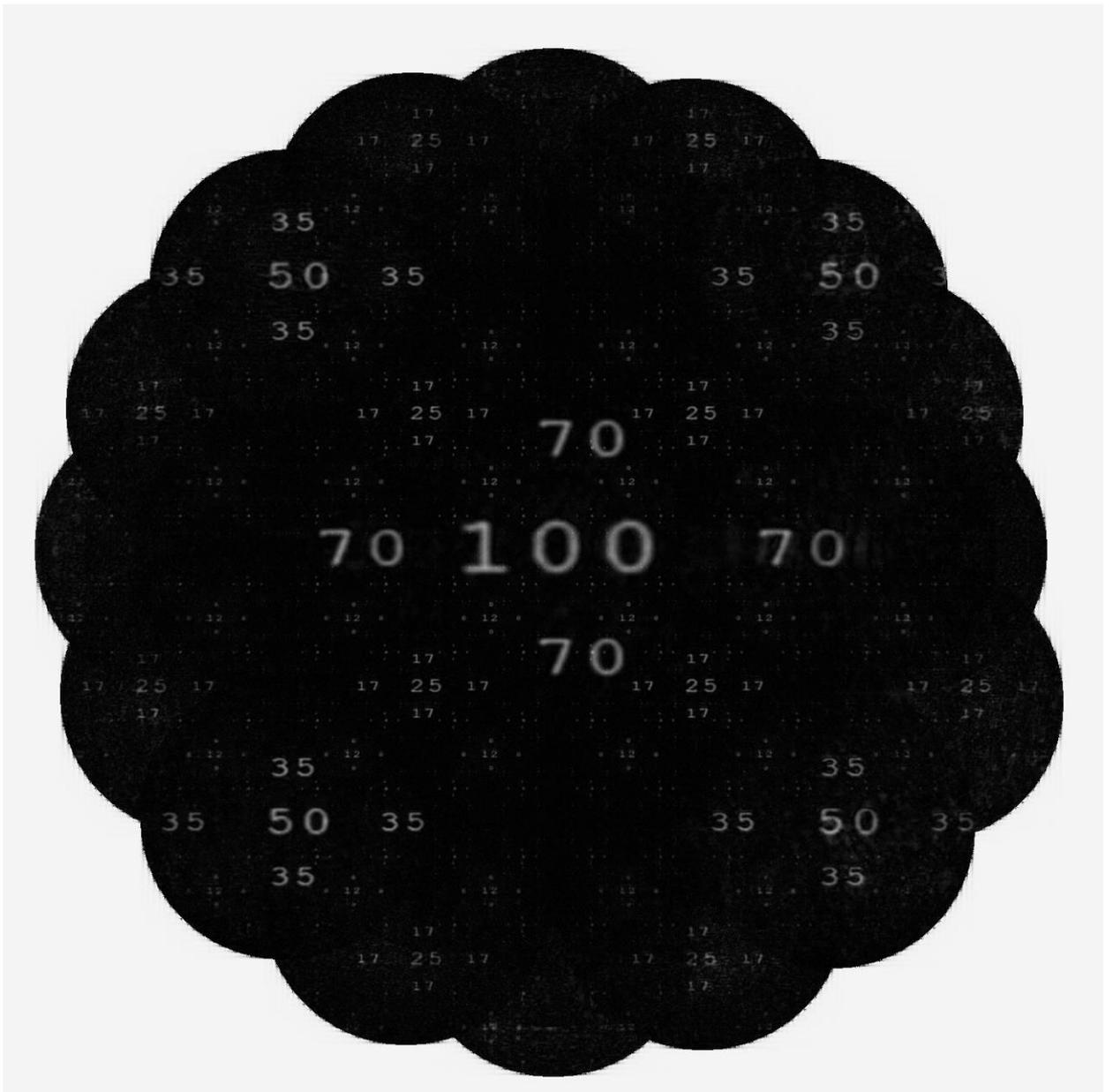

Fig.7. Reconstructed image after 10112 iterations (79 cycles for 128 difractograms). The size is 50x50 μ², white area at the edges is the remainder of $O_0(\vec{\tilde{\rho}})$ = 1.

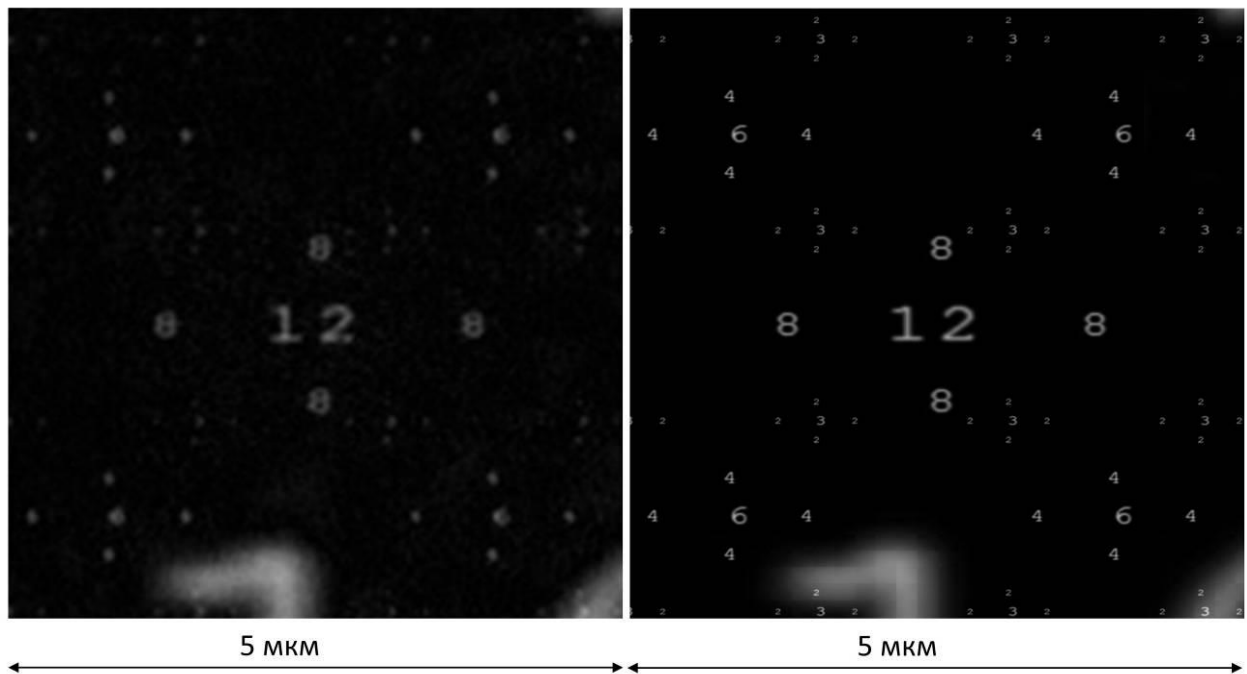

Fig.8. Comparison of the central regions of the reconstructed image (left) and the original one (right).

6. **Conclusion.**

In the past few years, ptychography has evolved into an all-wavelength microscopy method that does not require high resolution optical elements. It is widely used in X-ray, VUV and visible ranges. Thus, it may be safe to say that the idea of a lens-less optical system [1], which was proposed almost half a century ago, was finally practically realized. This has led to commercial ptychoscopes used in cytology [17-19]. The experimental setup for a lens-less microscope in its simplest form is comprised of four elements: a coherent light source, a platform with an object and a detector with a computer to analyze the difractograms (Fig.9). The current work serves as a brief introduction into the methods involved in ptychography. The wave packet in the far zone was chosen as the main diffraction integral, transferring the difractogram at the detector into the field at the object. This preserves the ability to acquire images with diffraction resolution ($\tilde{\delta} = \lambda/2$). At the same time, the computational algorithms still utilize the Fourier transforms of the fields at the object and the detector. Theoretically justified formulas to determine the size, domain discretization step at the object and spatial resolution were developed. Also, the distance between the object and the detector, the size of the latter as well as the size of the pixel are considered to be known. Requirements for the optimal dstance and detector pixel size were formulated for the square detector. The accuracy of the formulas used to determine the object domain was demonstrated using a point source as an example. Numerical ptychographic experiments in microscopy were conducted to demonstrate the possibility of acquiring the image of an object in the field of view 50x50 µ$^2$ with the resolution of 34 nm at the wavelength 10nm. The reconstruction time of a single object was approximately 1-2 hours on a personal computer, which can be further reduced by 2-3 orders of magnitude through software and hardware means. Further development in the ptychography field requires a broadening of its field of use as well as an experimentally-based systemic analysis of the accuracy and stability of the reconstruction algorithms.

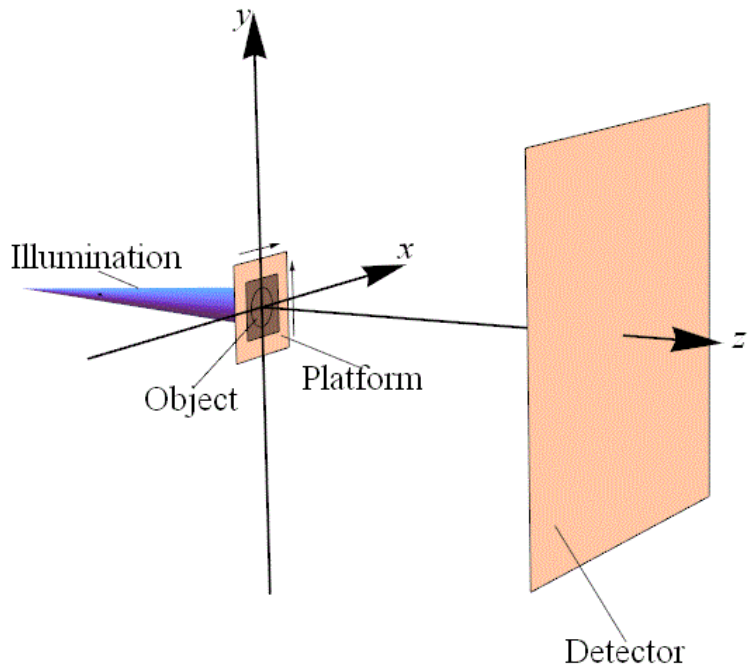

Fig.9. Ptychoscope layout.

## 7. Acknowledgements.

This work was supported by the Russian Federation Basic Funding within the project № 0023-0002-2018, and Russian Fund for Basic Research grants 19-02-00394, 18-08-01066, 17-08-01286, as well as by the Program for Scientific Research by the presidium of Russian Academy of Sciences «Current problems of photonics, probing of inhomogeneous media and materials» (PP RAN № 7).

The authors are thankful to A.V. Protopopov for the help in preparation of the manuscript.